# Comparison of H$_2$ and He carbon cleaning mechanisms in extreme ultraviolet induced and surface wave discharge plasmas

A Dolgov[1], D Lopaev[2], T Rachimova[2], A Kovalev[2], A Vasil'eva[2], C J Lee[1], V M Krivtsun[3], O Yakushev[3] and F Bijkerk[1]

[1] Dutch Institute for Fundamental Energy Research (DIFFER), Nieuwegein, The Netherlands
[2] Moscow State University, Moscow, Russian Federation
[3] Institute for Spectroscopy, Moscow, Russian Federation

E-mail: a.dolgov@differ.nl

**Abstract.** Cleaning of contamination of optical surfaces by amorphous carbon (a-C) is highly relevant for extreme ultraviolet (EUV) lithography. We have studied the mechanisms for a-C removal from a Si surface. By comparing a-C removal in a surface wave discharge (SWD) plasma and an EUV-induced plasma, the cleaning mechanisms for hydrogen and helium gas environments were determined. The C-atom removal per incident ion was estimated for different sample bias voltages and ion fluxes. It was found that H$_2$ plasmas generally had higher cleaning rates than He plasmas: up to seven times higher for more negatively biased samples in EUV induced plasma. Moreover, for H$_2$, EUV induced plasma was found to be 2-3 times more efficient at removing carbon than the SWD plasma. It was observed carbon removal during exposure to He is due to physical sputtering by He$^+$ ions. In H$_2$, on the other hand, the increase in carbon removal rates is due to chemical sputtering. This is a new C cleaning mechanism for EUV-induced plasma, which we call "EUV-reactive ion sputtering."

# 1. Introduction

According to the International Technology Roadmap for Semiconductors (ITRS), extreme ultraviolet lithography (EUVL) is currently the most advanced technology for the fabrication of integrated circuits with characteristic half-pitch, hp ≤ 22 nm [1]. Multilayer mirrors (MLM) are the basic optical element in EUV lithography. At an operating wavelength of 13.5 nm, MLMs consist of approximately 50-60 bi-layers of Mo:Si that are 6.7 nm thick. The mirrors are often covered with a protective layer that is 1.5-2 nm thick. To obtain the desired optical resolution, the mirrors must have a surface roughness much less than the wavelength, with reported values being approximately 0.2-0.3 nm. Thus, multilayer mirrors in EUV lithography are expensive, high-technology items, making it desirable to extend their useful lifetime as much as possible.

Previous research has demonstrated that MLMs lose their reflectivity due to the pollution of their surfaces with amorphous carbon, and/or surface oxidation, induced by intense EUV radiation [2-5]. In order to ensure optimal image resolution and uniform exposure, the EUVL optical system consists of 6-10 multilayer mirrors. In these circumstances, even a small loss of reflectivity for each mirror simultaneously (~1-2%) leads to a significant deterioration in optical throughput. For example [6,7], a carbon film just a few nm thick would already cause such loss of MLM reflectivity. In UHV conditions, where the rate of accumulation of amorphous carbon is limited by the partial pressure of hydrocarbons, the growth rate of the carbon layer can be quite high, ranging from 0.001 nm/hr to 0.01 nm/hr, depending on the precise EUV illumination and partial pressures [8.9].

Given that EUVL requires a long MLM lifetime (i.e. ~30000 hr) [10,11], it is obvious that cleaning mechanisms for MLMs are required. Moreover, such cleaning systems should be realized in such a way that interruptions to operation are minimized, i.e. *in situ*, without affecting the rest of the equipment or the lithographic process. In terms of quantities, such *in situ* cleaning should remove carbon with sufficient efficiency (at least faster than it deposits) and, at the same time, the capping layer of the mirror should not suffer damage over the entire period of the MLM's life.

At present, atomic hydrogen is used to clean MLMs [12-18], because the majority of basic hydrogen compounds are volatile. However, the efficiency of atomic hydrogen cleaning of amorphous carbon is extremely low [19] particularly in comparison with the surface recombination probabilities [20]. Moreover, atomic hydrogen generates heat (e.g., radiative heating from a hot filament), creating a high radiation and thermal load on the MLM surface that affects the speed and selectivity of cleaning. The efficiency of removing amorphous carbon from the MLM surface using plasmas is much higher than atomic hydrogen cleaning [19]. A low-temperature low-pressure plasma, even at low densities, provides fast, efficient and accurate carbon cleaning. This is why the generation and management of such plasmas, near the MLM surface during EUVL operation, is of special importance for the development of continuous *in situ* mirror cleaning technology. The EUV-photon energy of 91.8 eV is sufficient to generate a cascade of photons and secondary electrons. The EUV-induced secondary electrons collide with the background gas (mostly molecular hydrogen) to form a plasma above the surface of the EUV optical element. Furthermore, intensive EUV radiation, along with the flux of charged particles from the plasma, can induce processes on the surface with the participation of gas molecules.

This paper investigates the mechanism of amorphous carbon cleaning in EUV-induced plasma. To exclude surface sputtering and separate the influence of physical and chemical processes, the background gases were limited to molecular hydrogen and helium. Moreover, to determine the role of photons, analogous experiments were conducted in a low-temperature discharge plasma with similar ion energies and plasma densities.

By comparing cleaning rates between these different regimes, the roles of ion energy, EUV radiation, and chemical activity were clarified.

## 2. Samples and diagnostics of amorphous carbon

Previous research has shown that magnetron sputtered carbon films are structurally and chemically similar to carbon films that grow during EUV illumination [21]. Hence, in this work, magnetron deposited amorphous carbon films are used as model contamination layers. Films of carbon were deposited on silicon wafers and model MLMs (40 bi-layers of Mo/Si, each ~6.9 nm thick). The as-deposited carbon layers had thicknesses of ~20 and ~10 nm for the silicon wafer and MLM samples, respectively. Amorphous carbon layer thicknesses were measured using Raman spectroscopy (RS) and the Energy Dispersive Spectroscopy (EDS), i.e. X-ray fluorescence analysis (XRF). Figure 1 shows a typical calibration curve for RS and XRF. The symbols show intensity of the EDS and RS signals (carbon peak intensity for the EDS method and integral of the D- and G-peak region in RS) versus carbon film thickness. The deposited carbon mass was measured during deposition by a quartz mass balance. The layer thickness was measured using spectroscopic ellipsometry after deposition. A linear regression of the data results in a measurement accuracy of ±0.5 nm (whatever a monolayer is) for XRF and ±2 nm for RS. Since XRF provides higher precision, it was used for the data presented.

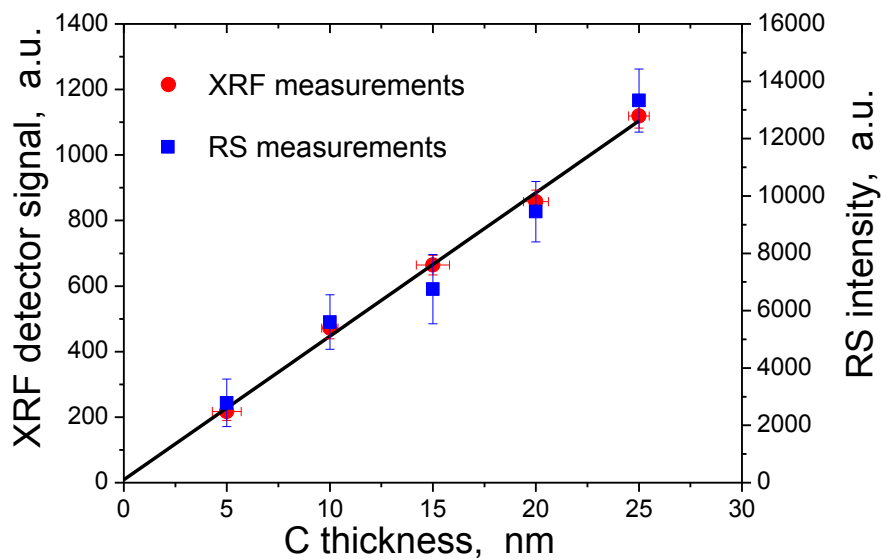

**Figure 1.** The magnetron-deposited carbon peak intensities for XRF (EDS) and RS versus the carbon thickness, measured by ellipsometry and quartz mass balance (during the deposition).

## 3. EUV-induced plasma

A schematic of the EUV-induced plasma interaction chamber and sample surface is presented in figure 2. The interaction chamber is capable of operating at background pressures of ~(3-5)·$10^{-8}$ Torr, which is maintained by separating the interaction chamber from the EUV source with a Zr/Si spectral purity filter (~40% transmission at 13.5 nm). The interaction chamber can be supplied with either $H_2$ or He, which are passed though a cold-trap to remove contaminants before the gas is introduced to the chamber.

The plasma is initiated by a short impulse of EUV 13.5 nm radiation, incident on the sample surface. The EUV source is a Sn-based, discharge Z-pinch, which operation principle has been described in detail elsewhere [22]. The source emits 100 ns pulses of broadband EUV radiation at a repetition rate of 1.5 kHz. After focusing and filtering, the spectrum is centered on 13.5 nm.

An image of the intensity profile of the EUV radiation at the focus is presented in figure 3a. The collector mirror's focal spot and a halo of scattered EUV radiation of 1-2 mm, due mainly to the roughness of the mirror surface are clearly visible. To eliminate most of the scattered EUV radiation, a diaphragm of synthetic mica (Ø = 8 mm) was placed on top of the samples (see figure 3a). Figure 3b presents a radial profile of the EUV radiation intensity inside the diagraph. It can be seen that the intensity of the EUV radiation reduces significantly towards the edge of the exposure zone, while the central area of 3 mm diameter has a relatively uniform distribution. The intensity profile of the EUV radiation incident on the sample was taken into account in the analysis presented below. The pulse energy was limited by the diaphragm to 0.042 mJ, whereas the average power was estimated to be 0.13 W/cm$^2$.

As a consequence of the fixed EUV source geometry, a cylindrical biasing electrode system was used. The sample plays the role of the cathode, being held at negative biases, down to -200 V. A metal cylinder, 30 mm in diameter and ~60 mm long, served as a grounded anode, with the sample centered on the axis of the cylinder, as shown in figure 2. In background-only tests, the photoelectron current has a duration that is nearly identical to that of the EUV pulse duration, while the peak current depends on the applied bias voltage (see figure 4a). In the presence of higher pressures (5-45 Pa), a similar photoelectron current pulse was observed. However, the tail of the pulse was found to extend to 4-10 μs (see figure 4a). Moreover, the current in the tail was found to increase for more negative bias voltages. The tail is a product of ionization by accelerated 'hot' secondary photoelectrons, as well as photoionization of gas molecules and atoms. However, it should be noted that the first process is, apparently, the main one, based on the known $H_2$ and He photoionization cross-sections, and the low pressures used in our experiments. The total charge developed by the EUV pulse (i.e. the number of electrons freed by ionization as well as photoelectrons) is estimated by integrating the current.

Figure 4b shows the total charge collected following the EUV pulse versus the applied bias voltage (volt-coulomb characteristics) in vacuum and at different values of $H_2$ pressure. The difference between sample currents in a gas and vacuum corresponds to the charge generated in the gas, i.e. plasma charge, and is related to the total ion charge collected by the sample surface. Assuming that the plasma is quasi-neutral, the total ion charge was used to estimate the number of ions formed in the volume of the EUV-induced plasma above the sample. It was also used to estimate the per-pulse ion flux, incident on the surface, for each applied bias voltage.

The radial profile of removed carbon approximately matches the profile of EUV radiation (see figure 3b). Figure 5 presents the rate of carbon removal at the center of the EUV spot in 3 Pa $H_2$ and He plasmas for various bias voltages. It can be seen that the rate of carbon removal increases with increasing bias voltage. Moreover, this rate is significantly higher for $H_2$ plasmas, compared to that for He plasmas. The increase in bias voltage leads to an increase in current and the energy of ions incident on the sample surface, which, should, naturally, lead to an increase in the rate of carbon removal. Note that the differences between the volt-coulomb characteristics for $H_2$ and He EUV-induced plasma are insignificant, i.e. the difference between ionic flux for $H_2$ and He is insignificant. This means that the number and kinetic energy of the particles incident on the sample surface for the two different plasmas are the same. Similarly, $He^+$ and $H_3^+$ particles have nearly identical momentum, so they should be equally effective at physical sputtering. Given these considerations, it is only possible to explain the great difference in the rate of carbon removal between hydrogen and helium by chemical processes, induced by hydrogen ions on the carbon surface. Such processes, as well as the role of EUV photons, are examined in detail in Section 5.

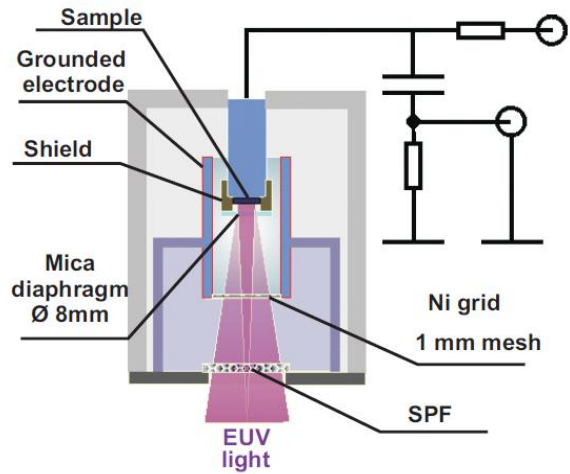

**Figure 2.** Set up of the experiment for carbon cleaning in EUV-induced plasma.

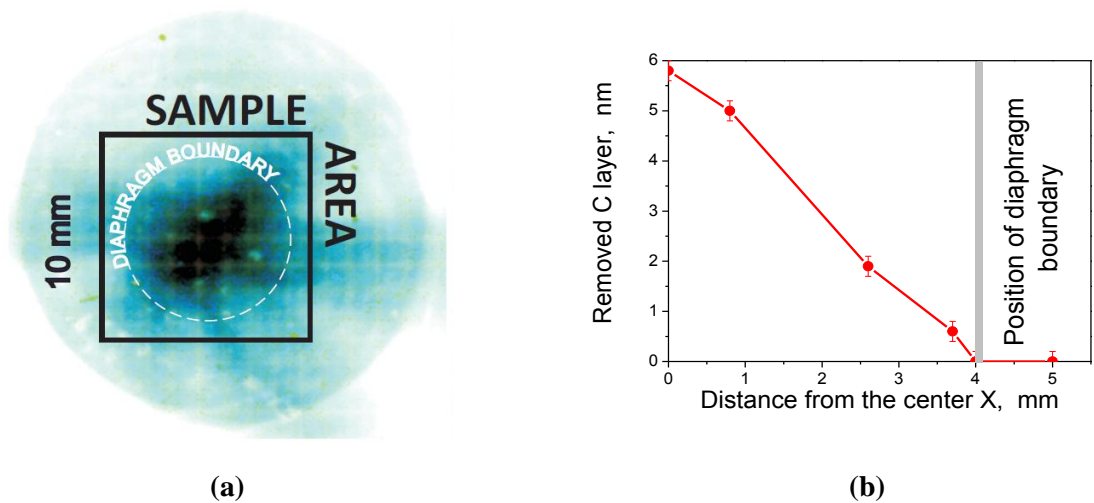

(a)                                                                   (b)

**Figure 3.** (a) EUV exposure pattern measured by sensitive foil and (b) C cleaning profile. To show (b) cleaning rate profile for obtaining the cleaning rate averaged on the profile to recalculate it then into the C yield atoms/ion since the total number of ions (charge of ions) came to the sample surface (on 1cm$^2$ per a EUV pulse) is measured as averaged on the sample area (inside the mica diaphragm, the position and size of which is indicated in yellow).

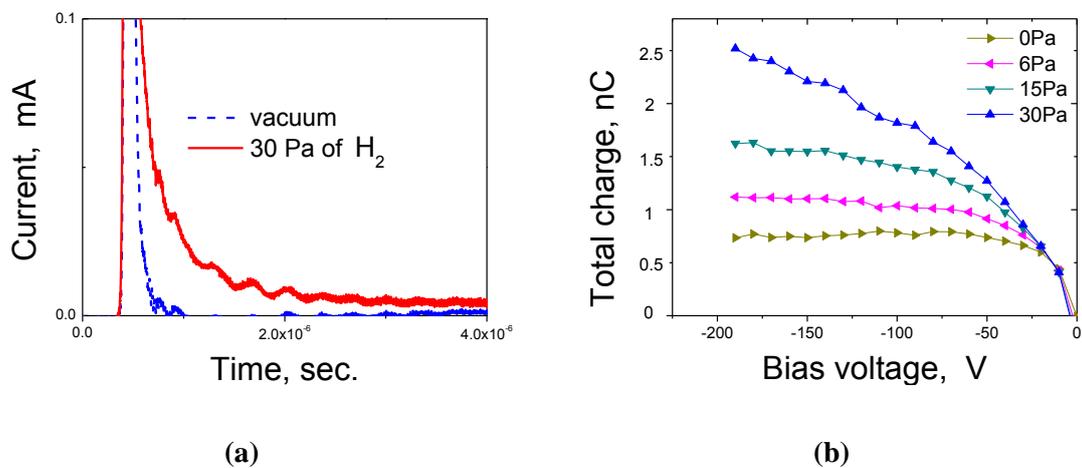

**(a)** **(b)**

**Figure 4.** (a) Time evolution of sample current in vacuum and gas (used to characterize the plasma formation). The appearance a slow tail in the sample current is clearly visible for an EUV pulse in the presence of a gas. The slow tail corresponds to the plasma decay current. The decay time is determined by slow ion motion in the plasma over the sample surface, where ions are accelerated by the plasma sheath in front of the sample surface, before they are incident on the sample surface. (b) Integrated CV characteristics for EUV diode. The total charge (the integral of the current curve, see figure 6) incident on a sample covered by carbon film vs the applied bias voltage at the different $H_2$ pressures.

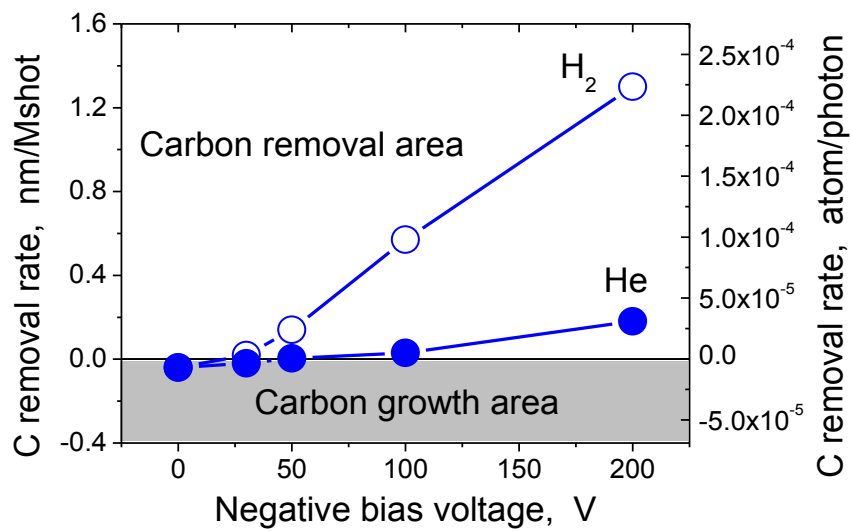

**Figure 5.** The averaged thickness of removed carbon in EUV-induced $H_2$ and He plasma vs the applied negative bias. The chamber pressure was 3 Pa, the EUV pulse energy was 85 μJ (~5·$10^{12}$ ph/$cm^2$pulse), and EUV dose was $10^7$ pulses. Negative values in the removed carbon thickness correspond to the growth of the carbon layer under the EUV exposure.

## 4. Low-pressure SWD discharge plasma

Even the simplest assessments, based on discharge current, voltage, and charge characteristics, demonstrate that the parameters of the plasma induced by EUV radiation are close to the parameters of a standard low-pressure discharge plasma, e.g. those used in microelectronics for the purpose of surface etching and cleaning. Thus, the removal of carbon due to EUV-induced plasma was compared to that of a low temperature discharge plasma to clarify the mechanisms of amorphous carbon removal.

The experimental setup is shown in figure 6. The plasma is formed in a long quartz tube (~100 cm long, 5.6 cm inner diameter) by means of a surface-wave discharge (SWD) at 81 MHz. RF excitation was achieved by two circular electrodes, placed close to one end of the tube. He or $H_2$ was supplied to the tube end at a pressure 2.7 Pa. The power supplied to the discharge was between 20 to 50 W, chosen to ensure that the plasma column extended from the electrode to the gas inlet, which served as a counter electrode for diagnostic measurements. The samples, which also served as an electrode to probe the plasma, were < 10x10 mm$^2$ in size, less than $10^{-3}$ of the counter electrode area.

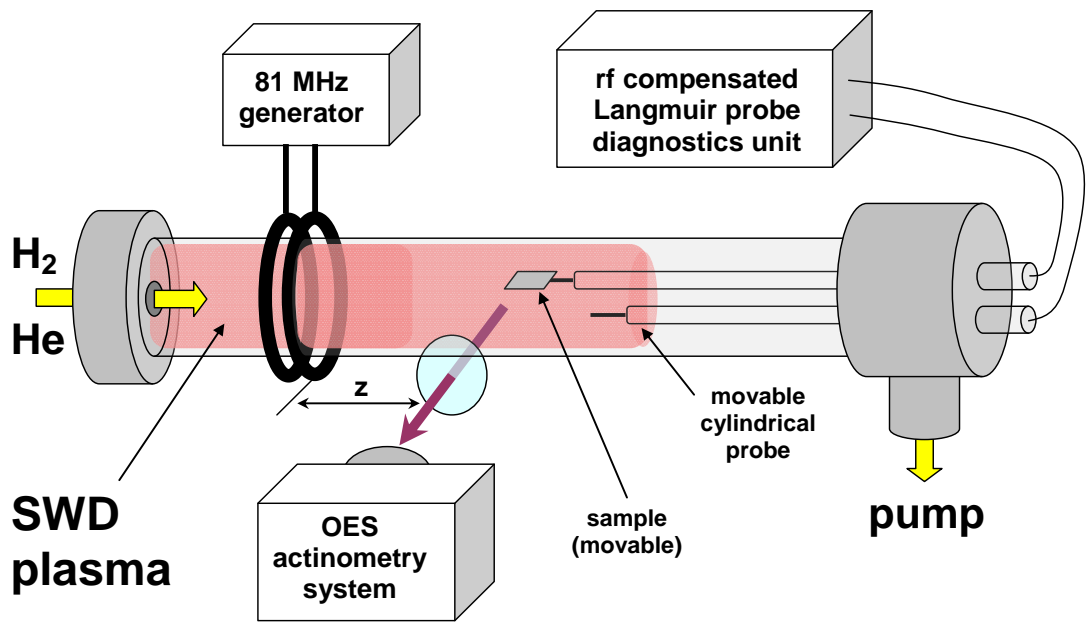

**Figure 6.** Set up for the experiments on C cleaning in SWD plasma, which has conditions similar to the conditions of EUV-induced plasma.

As is well known [23], the characteristic feature of the plasma column in surface-wave discharge is that the electron temperature and, accordingly, plasma potential and ion energy are nearly constant along the column. On the other hand, the plasma density and, accordingly, the ion flux incident on a surface, placed in the plasma, reduces with distance from the rf antenna. Therefore, the sample was placed on a travelling flat probe, able to move along the plasma column, thus, allowing the incident ion flux to be varied independently of the ion energy. To control the energy of the incident ions, the sample was biased with respect to the plasma potential. The sample was connected to a low-pass filter to remove the current induced by the plasma power supply. The parameters of the plasma above the sample surface were determined, based on measured sample voltage-current characteristics. Figure 7 shows typical VI characteristics of the sample in He and $H_2$ at different distances from the rf antenna.

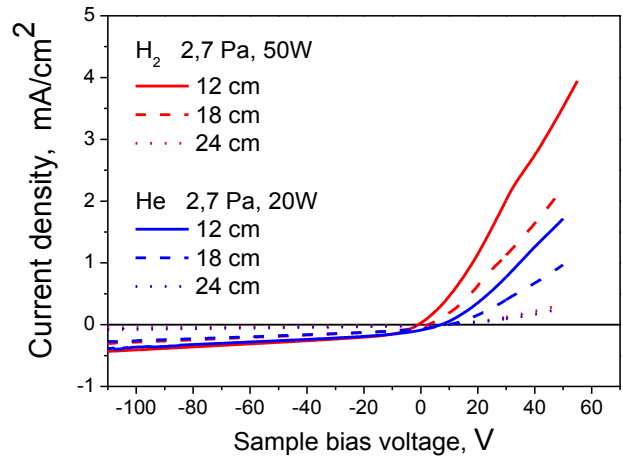

**Figure 7.** VI characteristics for the flat sample in SWD plasma column in $H_2$ (solid lines) and He (dashed lines) at the different distances, *z*, from the rf antenna.

To verify the consistency of the measurements of the plasma parameters using the sample as a flat probe (and, as a result, the energy of ions on the sample surface), the experiment also used an rf-compensated standard cylindrical probe (~10 mm long, ~0.1 mm in diameter). It should be noted that, compared with the "small" cylindrical probe, some distortion is observed in the electron part of the VI characteristics of the sample – the well-known electron depletion effect – that is quite expected for large-area probes. Therefore, the sample measurements are expected to underestimate the plasma density, which is extracted from the electron part of VI characteristics, while accurately estimating the electron temperature, $T_e$. At the same time, the plasma densities extracted from the ion part of VI characteristics of the cylindrical and the sample appear to be in a good agreement, although the definition of a "flat probe" for the sample cannot be applied exactly in this case because of the low plasma density. As was observed, the plasma sheath over the sample (probe) is rather big and not purely flat. Taking into account that this fact can lead only to small discrepancies, a "flat-probe" approach was used for calculating the energy and ion flux at the sample surface.

As an example, figure 8 demonstrates the rate of carbon removal in hydrogen plasma (black symbols and lines) and helium plasma (grey symbols and lines) versus ion flux for different bias voltages ($U_{bias}$=0 V and $U_{bias}$=-90 V). With an estimated density of amorphous carbon film of ~1.9 $g/cm^3$ [24], the rate of carbon removal was calculated in atom/($cm^2$s) on the right scale. Open and filled symbols in figure 8 correspond to XRF (EDS) and RS data, respectively. The trend lines are the linear fits to the data. The slope of each line is interpreted as ***carbon removal probability*** or ***C yield*** per ion for the respective conditions (bias voltage and gas species).

The C yield for both $H_2$ and He plasma is shown in figure 9 as function of the negative bias applied to the sample. The error bars on vertical axis correspond, mainly, to the uncertainties in the measurements and calculations of the C cleaning rate. As can be is seen in figure 9, the C yield is higher for $H_2$ plasma than for He plasma, and this difference increases with increasing bias. It should be noted that this plot also differs from the similar plot for EUV-induced plasma (compare with figure 5). It shows that, while the average plasma conditions are similar, the characteristics of EUV-induced plasma are dominated by the time-varying flux and intensity of ionizing radiation, incident on the carbon surface. A comparative analysis of the carbon removal mechanism in discharge plasma and EUV-induced plasma allows the role of the EUV radiation to be understood.

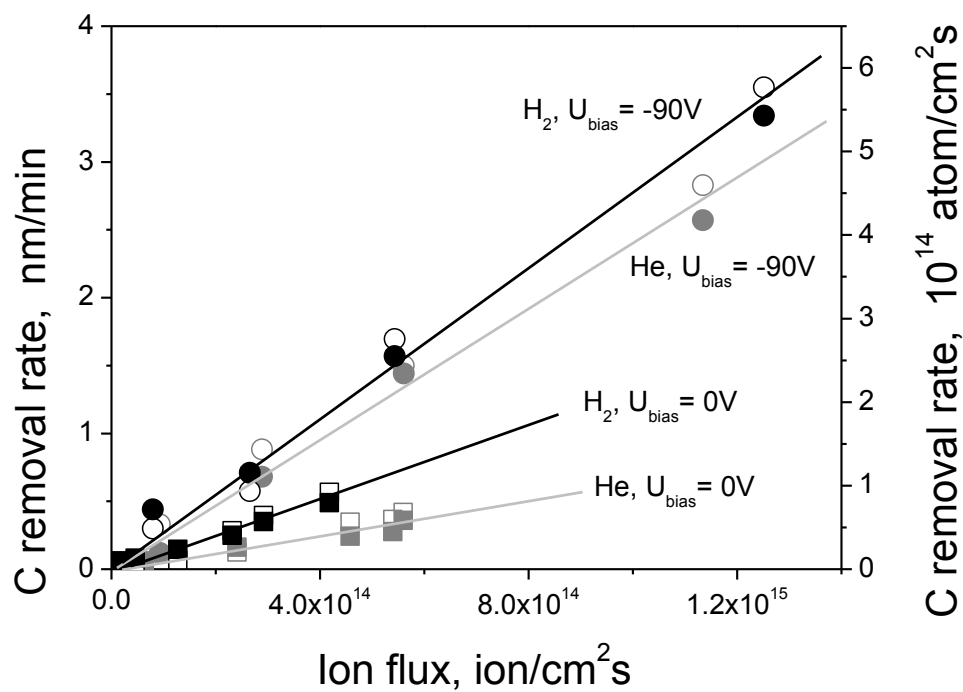

**Figure 8.** Carbon cleaning rate vs the ion flux (i.e. $H_3^+$ and $He^+$ for hydrogen and helium plasmas, respectively) at two (-90 V and 0 V) biases ($U_{bias}$), applied to the sample. Data for hydrogen and helium are shown by black and grey colors, respectively. The lines are linear fits to the data. Open and filled symbols correspond, respectively, to XRF (EDS) and RS measurements of the carbon film thickness.

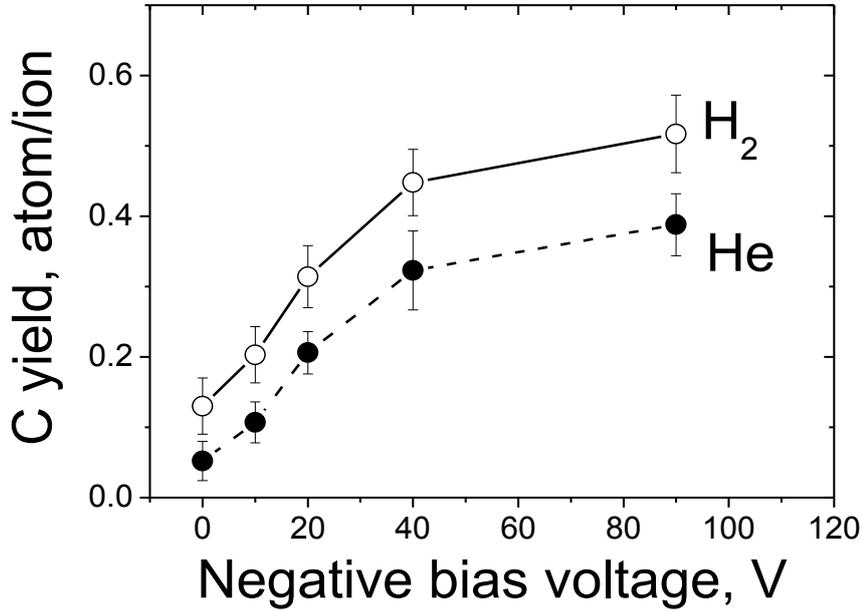

**Figure 9.** The C yield (atom/ion) versus the negative bias applied to the sample in $H_2$ (open circles) and He (filled circles) SWD plasmas.

## 5. Discussion

To understand the detailed mechanism of C cleaning in an EUV-induced plasma and to characterize the role of EUV photons in this process, the EUV-induced and discharge plasmas should be compared in a physically clear manner. The role of EUV radiation in C-cleaning can be understood by comparing C-cleaning measurements with and without EUV radiation, but with similar fluxes and energy of the incident ions. For this purpose, we will try to find a relation between the ion energy, the applied bias, and plasma parameters in both types of plasmas.

The energy spectrum of the ions incident on the sample surface from the almost collision-less plasma sheath should have a notable peak. The peak of the ion energy spectrum and the peak width can be estimated as $\sim U_{bias} + U_p$ and $\sim 2U_f$, respectively, where $U_p$ and $U_f$ are plasma and float potentials. In reality, the peak of the ion energy spectrum is somewhat lower, and the ion energy distribution is wider than given by this simple estimation. However, the validity of such a treatment has been confirmed by Monte Carlo calculations of the ion energy spectrum above sample surface [25]. Model calculations and estimates show that, because of the fast reaction: $H_2^+ + H_2 \rightarrow H + H_3^+$ ($k > 10^{-9}$ cm$^3$/s for $T_i < 1$ eV in the plasma volume), and the given experimental conditions, ($H_2$ pressure 2.7 Pa, plasma density $\sim 10^9$ cm$^{-3}$) $H_3^+$ is the dominant ionized species. [26]).

In contrast to the discharge plasma, the type of ion and ion energy spectrum cannot be easily predicted using only qualitative considerations for the EUV-induced plasma. According to model calculations [25, 26], $H_3^+$ is still the dominant ion, but the energy distribution of both ions ($H_3^+$ and $H_2^+$) is noticeably wider than the energy distribution of the ions in the continuous SWD plasma. In addition, the peak of ion energy distribution is a factor of two lower compared to that of the SWD. case energy equal to 0,5 U bias.

In experiments with the discharge plasma, the carbon removal yield per ion is assessed directly from the data (see above). For the pulsed EUV-induced plasma, the yield is calculated as the average number of removed carbon atoms per EUV pulse divided by number of ions incident on the sample surface per pulse:

$$R_C = \frac{N_C}{N_i} = \frac{\rho \cdot S \cdot \langle h \rangle \cdot e}{m_C \cdot Q_i}$$

Where $\rho \approx 1.9$ g/cm$^3$, is the carbon density, $S$ is the area of the focus, $\langle h \rangle$ is a spatial average of the thickness of the removed carbon layer, estimated from the measured C layer thickness profile inside diaphragm. The charge is of the ion is $e$, $m_c$ is the mass of the carbon atom, and $Q_i$ is the total number of ions incident on the sample surface. $Q_i$ was determined by integrating the ion current.

C cleaning yields, calculated for both discharge and EUV-induced for hydrogen (a) and helium (b) plasmas, versus the peak ion energy, calculated as described above, are shown in figures 10a and 10b, respectively. As was mentioned above, the error bars shown on the horizontal axis represent the approximate width of the ion energy distributions for both plasmas, while the error bars on vertical scale correspond to the uncertainties associated with measurements of the C yield. For helium, the carbon yields for EUV-induced and SWD plasmas are rather similar, while for hydrogen, the two are quite significantly different. This shows that the mechanism of carbon removal in helium plasmas is the same for both SWD and EUV-induced plasmas, and the role of EUV radiation is insignificant. For the hydrogen plasma, the differences show that the EUV radiation is inducing additional processes that increase the cleaning rate beyond a simple combination of physical and chemical sputtering.

Given the inert nature of helium ions, the removal of carbon from the surface is most likely due to physical sputtering only. Physical sputtering has been studied in the literature for virtually all elements of the periodic table, in both neutral and ionized states. Sputtering (sputtering yield) depends on the mass, charge, and energy, $E_i$, of the incident ions, as well as the surface binding energy, $E_s$, of the target atoms, etc. An accurate model of physical sputtering and resulting sputtering yield calculations can be found in [27]. In this review, a semi-empirical expression for the sputtering yield for almost all combinations of incident ion and target material was obtained, which gives a limited set of free parameters to fit in order to achieve reasonable agreement between observed and calculated sputtering yields Bohdansky et al. [28] Yanamura et al. [29]. The sputtering yield for light ions is given by:

$$Y(E_i) \approx \frac{A}{E_s} \cdot \frac{S_n(\varepsilon)}{1+B\varepsilon^{0.3}} \cdot \left(1 - \left(\frac{E_{th}}{E_i}\right)^{1/2}\right)^{2.5} \quad (1)$$

where $A$, $B$ are empirical coefficients, for which we used values obtained by Yamamura and Tawara [29].. $E_{th}$ is the threshold energy, which is taken as a free parameter. $S_n(\varepsilon)$ is the nuclear stopping cross-section, which can be expressed in terms of the elastic cross-section, $s_n(\varepsilon)$, and can be found by substituting a reduced ion energy, $\varepsilon$, into the analytical expression:

$$s_n(\varepsilon) \approx \frac{3.441\sqrt{\varepsilon}\ln(\varepsilon+2.718)}{1+6.355\sqrt{\varepsilon}+\varepsilon(-1.708+6.882\sqrt{\varepsilon})} \quad (2)$$

The values of the parameters in equation (1), obtained from ref [29] for carbon surface bombarded by $He^+$ and $H_3^+$ ions, are presented in table 1.

**Table 1.** The parameter values for equation 1 for $He^+$ - C and $H_3+$ - C sputtering according to [29].

|  | Physical ion sputtering ($He^+$) | Physical ion sputtering ($H_3^+$) |
|---|---|---|
| $A$ | 0.0342 | 0.0407 |
| $B$ | 0.335 | 0.357 |
| $S_n(\varepsilon)$ | $115.03 \cdot s_n(\varepsilon)$ | $49.05 \cdot s_n(\varepsilon)$ |
| $\varepsilon$ | $9.2 \cdot 10^{-5} \cdot E_i$ | $2.09 \cdot 10^{-4} \cdot E_i$ |
| $E_{th}$ | $3.87 \cdot E_s$ | $3.79 \cdot E_s$ |

As follows from expression (1), the most important parameter for the mechanism of physical sputtering is the binding energy of the target atoms: amorphous carbon in our case. The binding energy also determines the threshold energy, $E_{th}$. As an estimate, $E_s$, is usually assumed to be qual to the sublimation energy (enthalpy of vaporization $\Delta H_C$) of the target material, which, for graphite-like surfaces, is $\Delta H_C \approx 7.43$ eV. However, as shown [30,31] in the analysis of amorphous carbon sputtering by light ions, experimental data are well described using a smaller value of $E_s \approx 4.5$ eV.

Therefore the expression (1) was fit (dashed lines in figures 10a and 10b) to the experimental data for both SWD and EUV-induced plasmas, using the parameters from table 1, and taking $E_s$ and $E_{th}$ as free parameters. The values of $E_s$ and $E_{th}$ for $He^+$ (see table 2) were found to be even lower than those found in ref [30,31], which can be clearly seen in figure 10a and 10b when considering the C sputtering yield at low ion energies.

Rigorous methods, that directly model the interaction of ion beams with a surface, such as TRIM (TRansport of Ions in Matter) were also used to understand the carbon cleaning mechanism. TRIM is a Monte Carlo computer program that calculates the interactions of energetic ions with amorphous targets. The threshold energy, $E_{th}$, for sputtering amorphous materials is usually found to be close to the energy required to displace a target atom $E_{displ} = 25$ eV (see table 3). Although TRIM is not optimum for non-monochromatic ion-energy spectra, by approximating the ion-energy spectrum by its average value, the rate of ion sputtering of carbon films by $He^+$ and $H_3^+$ ions was estimated, and used to gain insight into the mechanism of the carbon removal. $H_3^+$ ions were approximated by two different methods: $H_3^+$ ions were assumed to be a hydrogen ion with a mass of 3 amu, and, in the second case, as a helium ion with a mass of 3 amu. The results for these two cases were found to differ by less than 1%. As above, the density of the carbon target is taken to be 1.9 g/cm$^3$. Three parameters: $E_{displ}$, $E_{bin.lat.}$ and $E_{bin.surt}$ were varied to obtain a best fit (solid lines) to the data in figures 10a and 10b. $E_{bin.lat.}$ is the binding energy of non-surface carbon atoms in the layer, while $E_{bin.surt}$ is the surface binding energy of carbon atoms. The values of the TRIM calculation parameters are presented in table 3. As can be seen from the disagreement between the best-fit values obtained from TRIM and those reported in the literature (see table 3), the removal of the amorphous carbon in an EUV-induced plasma cannot be described by kinetic sputtering. For the case of He, this is not surprising, since physical sputtering is the only available mechanism and the EUV pulse simply generates the plasma to initiate sputtering..

As follows from (1) and TRIM calculations, the removal of carbon in the discharge $H_2$ plasma has a similar character to the removal of C by $He^+$ ions, but the C yield is higher. However, compared to $He^+$ ions, agreement with experiment data can only be achieved with a substantial reduction of the threshold energy (see tables 2 and 3). For example, the TRIM calculation shows that the displacement energy is almost equal the surface binding energy. This is usually interpreted as the possibility of a direct or ion-stimulated reaction between implanted ions (atoms) with the displaced ("free") target atoms [32]. This mechanism of removal of target atoms by low-energy ions, commonly called chemical sputtering, is well known for H sputtering of carbon [30-32].

**Table 2.** Parameters in fitting (1) used for the three data sets in figures 10a and 10b.

|  | Physical ion sputtering ($He^+$) | Chemical ion sputtering ($H_3^+$) | RIE (EUV) + chemical ion sputtering ($H_3^+$) |
|---|---|---|---|
| $E_{th}$ (eV) | 7.35 | 2.27 | 1.52 |
| $E_s$ (eV) | 1.9 | 0.6 | 0.4 |

**Table 3.** TRIM model parameters used for the three data sets in figures 10a and 10b.

|  | Physical ion sputtering ($He^+$) | Chemical ion sputtering ($H_3^+$) | Recommended by TRIM(graphite) | RIE (EUV) + chemical ion sputtering ($H_3^+$) |
|---|---|---|---|---|
| $E_{displ.}$ | 15 | 4.5 | 25 | 4.5 |
| $E_{bin.lat.}$ (eV) | 2 | 2 | 3 | 2 |
| $E_{bin.surf.}$ (eV) | 4.5 | 3 | 7.37 | 1.4 |

The penetration depth of low-energy hydrogen ions (with energies of tens of eV) in carbon is small enough ($\leq 3$ nm) to allow the resulting volatile species to diffuse to the surface and escape (mainly $H_2$ and $CH_4$) [33].

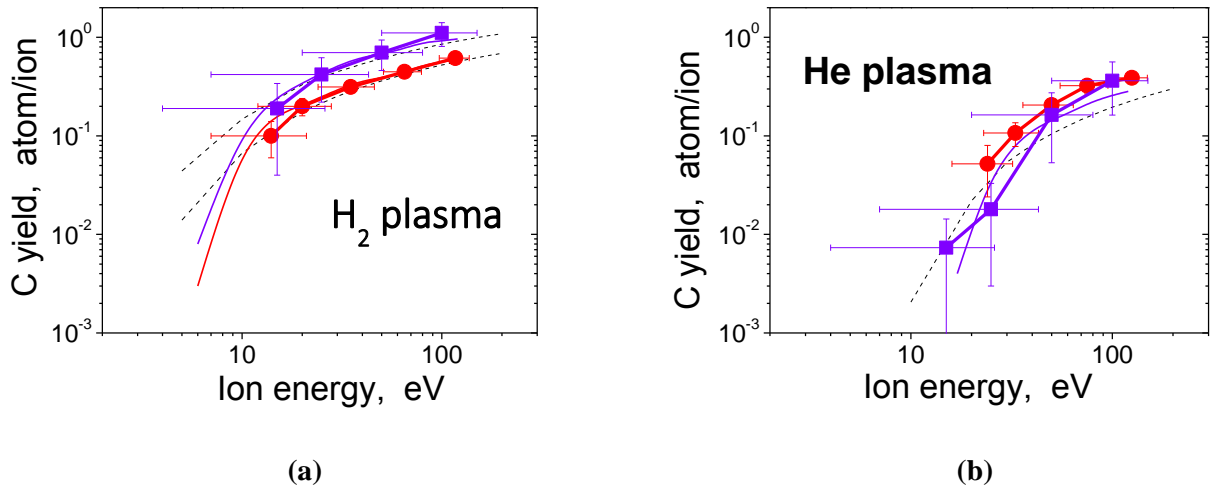

(a)    (b)

**Figure 10.** C atom yield (atom/ion) vs the mean energy of ions incident on the amorphous carbon surface in $H_2$ (a) and He (b) plasmas. Big dark circles with thick lines between them are experimental data for SWD plasma while big dark squares with thick lines between them are experimental data for EUV-induced plasma. Dotted black lines are fits according to expression (1) with parameters from table 1 and 2. Small symbols and solid thin lines are estimates made for $H_3^+$ and $He^+$ ions by using TRIM calculations with parameters presented in table 3.

Atomic hydrogen, formed in plasma due to gas dissociation, may promote carbon removal. However, the flux of atomic hydrogen, incident on the sample surface, should be small in both the EUV-induced and SWD plasmas. For example, the estimates of the density of H atoms in the discharge tube, from actinometry on argon atoms [20], showed that the relative concentration of atomic hydrogen was ~0.5-1%. According to the results of [30], and the measured ratio of hydrogen ionic and atomic fluxes, incident on the sample is such that the rate of the reactive ion etching (RIE) involving hydrogen atoms should not exceed the rate of chemical sputtering for SWD hydrogen plasmas.

Indeed, in the EUV-induced hydrogen plasma, the C atoms yield per ion is higher than in the SWD plasma. To obtain agreement between experimental results and calculations using only physical sputtering, the energy threshold has to be made very small (much smaller than any physical arguments would allow). Additionally, it should be expected that the energy threshold is independent of the plasma generation mechanism. The difference between the EUV-induced and the discharge plasmas in $H_2$ is relatively constant for all ion energies, which offers indirect evidence that this difference is due to the EUV pulse, since the EUV photon fluence is the same for all data plotted in figures 10a and 10b. This increase in the yield of carbon for EUV-induced plasma can be explained only by an additional *chemical* mechanism, i.e. removal by the additionally formed atomic hydrogen.

This is also indirectly confirmed by TRIM calculations, in which, to describe the carbon yield in the EUV-induced hydrogen plasma, $E_{bin.surf}$ has to be reduced practically to an unrealistically low value (see table 2). According to [30,31] this can only be due to ion-stimulated reactions between hydrogen and carbon on the surface. This process is analogous to the process of RIE [30] with the difference that the atomic hydrogen is formed directly on the surface layer of carbon, due to the dissociation and ionization of adsorbed molecular hydrogen, followed by the dissociated species being "dissolved in the surface layer." Furthermore, the EUV photons that are absorbed in the carbon generate excited species (carbon radicals and ions). At the carbon layer surface, we estimate from reflection loss data and secondary electron yield data [34] that up to 4% of the surface carbon atoms will be either ionized or form a radical. We will refer to this combination of processes as reactive ion sputtering (RIS).

In a sense, the process of the formation of atomic hydrogen on the surface is similar to the process of surface contamination and oxidation of surfaces. As is known, these processes are caused by dissociation of hydrocarbons and water during exposure to EUV radiation [2-7]. Moreover, the oxidation of carbon contamination under EUV radiation is even observed in ultra-high vacuum.

To estimate if surface dissociation is significant, the surface coverage of molecular hydrogen must be estimated. The energy of desorption and residence time for hydrocarbon molecules and hydrogen molecules are very different. For polyatomic hydrocarbon molecules, the desorption energy is much higher than that of hydrogen. Therefore, even with large concentration differences in the volume above the sample, the surface concentrations of hydrogen and hydrocarbons may be comparable. These considerations allow the desorption energy of $H_2$ from an amorphous carbon surface during EUV illumination to be estimated.

Assuming that the increase in the carbon yield in the EUV-induced plasma is due only to the EUV-dissociated hydrogen (i.e. almost half of the carbon is removed by RIS), the removal of one C atom is estimated to require approximately $10^3$ EUV photons. If it is also assumed that RIS results in $CH_4$, the minimum fraction of the surface occupied by dissociated $H_2$ molecules can be estimated to be $\theta \approx 2 \cdot 10^{-3}$. Sorption equilibrium coverage is given by:

$$H_2 + surface \underset{v_d}{\overset{k}{\leftrightarrow}} H_2^{surface} \qquad \theta = \frac{k[H_2]}{v_0 e^{-\frac{E_d}{kT}}} \qquad (3)$$

(where $[H_2] \sim 7 \cdot 10^{14}$ (3 Pa), $k \sim 10^{-10}$ cm$^3$/s is the rate constant corresponding to the sticking coefficient ~1 and, in fact, is the rate constant of collisions in gas phase, $v_0 \sim 10^{13}$ s$^{-1}$ is the so-called "attempt frequency", that can be interpreted as the vibrational frequency of the adsorbed species in a potential well with a depth corresponding to that of the desorption energy, $E_d$, so that only the fraction of the particles having an energy higher than $E_d$. From eq (3), $E_d \sim 380$ K for surface adsorbed atomic hydrogen (ignoring atomic hydrogen "dissolved" in the surface layer). Nevertheless this estimate corresponds to the desorption energy of molecular hydrogen measured on different materials, both metals [20] and dielectrics [35]. It should be noted that the density of adsorption sites is correlated with surface defects, rather than simply the density of surface atoms. In this sense, the estimated value of $E_d$ characterizes not only the surface but also the depth of $H_2$ diffusion [33]. Therefore, it reflects a certain characteristic binding energy of $H_2$ molecules with the surface layer of carbon, part of which is activated by EUV radiation and the plasma. This "activated" (chemically active) carbon in principle is able to react with $H_2$, while atomic hydrogen is able to react with carbon whether it is activated or not.

Thus, under EUV radiation, carbon removal is not only due to physical and chemical sputtering by ionized species, but is also due to the creation of volatile species from reactions between radicals and between radicals and molecular species. Since radicals are not influenced by bias voltages, there is very little atomic hydrogen incident on the surface. Instead, the radicals are created by the dissociation of both the carbon surface, and adsorbed molecular hydrogen.

## 6. Conclusion

In the given work, the removal of thin amorphous carbon films from Si substrates in two types of plasmas in presence of two types of gas: helium and hydrogen, was studied. The experimental data clearly show that the measured carbon removal rates for SWD and EUV-induced He plasmas are similar. He removes carbon through physical sputtering. Hydrogen was found to work differently. It was shown that the hydrogen cleaning mechanism in the case of an EUV-induced plasma is about 5 times more efficient then in a SWD plasma. The qualitative analysis showed that the rate of cleaning in EUV-induced $H_2$ plasma is greater than expected owing to reactive ion etching by atomic hydrogen, adsorbed on the surface of carbon film. Analytical and computational models were used to show that the data do not support physical sputtering in the case of hydrogen plasmas. Yet, a consistent carbon binding energy and threshold energy for both SWD and EUV induced plasmas could be found once chemical sputtering is taken into account. However, an EUV plasma was found to create a larger ratio of atomic versus ionic hydrogen on the sample surface than in the case of SWD plasma.

**Acknowledgments**

This work is part of the research program "Controlling photon and plasma induced processes at EUV optical surfaces (CP3E)" of the "Stichting voor Fundamenteel Onderzoek der Materie (FOM)" which is financially supported by the Nederlandse Organisatie voor Wetenschappelijk Onderzoek (NWO). The CP3E programme is cofinanced by Carl Zeiss SMT GmbH (Oberkochen), ASML (Veldhoven), and the AgentschapNL through the Catrene EXEPT program.